\documentclass[preprint,aps,amsmath]{revtex4}
\usepackage{graphicx}

\begin{document} 

\title{ZM theory II: \\
Hamilton's and 
Lagrange's equations of motion}

\author{Yaneer Bar-Yam}

\affiliation{New England Complex Systems Institute \\ 24 Mt. Auburn St., Cambridge, Massachusetts 02138}

\begin{abstract}
We show that considering time measured by an observer to be a function of a cyclical field (an abstract version of a clock) is consistent with Hamilton's and Lagrange's equations of motion for a one dimensional space manifold. The derivation may provide a simple understanding of the conventions that are used in defining the relationship between independent and dependent variables in the Lagrangian and Hamiltonian formalisms. These derivations of the underlying principles of classical mechanics are steps on the way to discussions of physical laws and interactions in ZM theory. 
\end{abstract}


\maketitle

\section{Overview}   

In this paper we continue an investigation of the implications of a novel assumption about the relationship of fields, space and time.\cite{ZM1} The central assumption is that time can be obtained from the values of a field variable. This treatment of the field variable $Z$ over the space manifold $M$ leads to ZM theory. The theory will be developed through a sequence of papers that explore the implications of such an assumption considering the compatibility of the approach with conventional physical laws, principles, mechanics, elementary forces and excitations. This paper is largely devoted to showing that Hamilton's and Lagrange's equations of motion \cite{Goldstein,Sussman} are consistent with the assumptions for a one dimensional space. We continue the practice of being meticulous about providing details of derivations to clarify the assumptions.

\section{Introduction}   

We review the basic framing of ZM theory from the first paper in the series, ZM1 \cite{ZM1}  with sufficient detail for this paper to be essentially self-contained. We consider a system, $Z$, with some set of distinctly labeled states that perform sequential transitions in a cyclic pattern, i.e. an abstract clock. Discreteness of the clock will not enter into the discussion in this paper. The clock states can therefore be extended to cyclical continuum, $U(1)$. The state change of the clock defines proper time, $\tau$, as defined by the clock. Since the clock is cyclical it is possible to represent the changing state using an oscillator language: 
\begin{equation}
\psi  = \exp ( - im \tau ), 
\end{equation}
where 
\begin{equation}
m = 2\pi /T
\end{equation}
is the cycle rate in radians. The clock phase is $c = m \tau$ modulo $T$, though this expression is not analytic so that derivatives should be defined in terms of $\psi$. However, locally with proper choice of the location of the discontinuity, or where analytic continuation is valid, derivatives can be defined in terms of $c$. 
The notation is chosen anticipating that $m$ will become the `rest mass' of the clock when it is reinterpreted as a particle. 
{\em A-priori} there is no difference between clockwise and counter-clockwise rotation, however, once one is identified it is distinct from the other.

To introduce the space manifold, $M$, we consider a parameter, $x$, associated with the environment such that properties of the environment may lead to variation of the clock state with $x$ (we assume $x$ is a real number parameter, which, for convenience, is measured in the same units as $\tau$). The assumption of ZM theory is that observer time can be obtained by the rate of change of the clock, $c$, as a local gradient of the clock state in the two dimensional space given by $\hat \tau$ and $\hat x$ treated as a Euclidean space, whose direction can be considered as a `direction of time:'
\begin{equation}
(m, - k) =  m \hat \tau  - k\hat x.
\label{clock}
\end{equation}
We assume that time measures Euclidean distance along the time direction in the same units as $x$ and $\tau$, so the rate of change of the clock phase in the direction of time is 
\begin{equation}
\omega  = \sqrt {m^2  + k^2 }, 
\label{omega}
\end{equation}
i.e. consider possible directions of time  
\begin{equation}
\hat s =  \cos (\zeta )\hat \tau + \sin (\zeta )\hat x , 
\end{equation}
where $\zeta $ is the angle of rotation of $\hat \tau$ into $-\hat x$, with a clock change 
\begin{equation}
\begin{array}{ll}
d_s c &= \hat s \cdot (\partial _{\hat \tau}  c,\partial _{\hat x} c) \\
&= m \cos (\zeta ) + k\sin (\zeta ); 
\end{array}
\label{generaltime}
\end{equation}
maximizing this with respect to the angle $\zeta $ gives 
\begin{equation}
\tan (\zeta ) = k/m, 
\label{geometry}
\end{equation}
and 
\begin{equation}
\omega  = d_t c = \sqrt {m^2  + k^2 }, 
\label{omegatime}
\end{equation}
where $\omega$ is defined to be the magnitude of the rate of change of the clock phase.


We can specify locations in the two dimensional space by their coordinates along the axes 
\begin{equation}
(\tau_0,x_0)=\tau_0 \hat \tau + x_0 \hat x .  
\end{equation}
We can then write $t$ and $x$ in terms of $\tau_0$ and $x_0$. All expressions can be directly obtained from the geometry of the angle of time with space. For the case of uniform variation of clock time in space, and assuming time has no dependence on the observer defined space $x$, time is given by
\begin{equation}
t = (\omega/m) \tau_0.
\label{ttau00}
\end{equation}
The direction of time implies a shift in the position of the origin of $x$ toward negative values and a reference location translates to the right, with $x$ given by: 
\begin{equation}
\begin{array}{ll}
x&=x_0+(k/m)\tau_0 \\
&=x_0+(k/\omega) t.
\end{array}
\end{equation}
This identifies a connection between the velocity and $k$ in ZM theory. 

\section{Spatial variation and Hamilton's equations of motion}

The determination of the direction of time through extremum of the clock rate of change is reminiscent of variational principles, and indeed, we will show that it is consistent with Hamilton's and Lagrange's equations of motion, for a single dimension of space. To facilitate the connection we will use notation commonly adopted for advanced classical mechanics.\cite{Goldstein}
For the treatment of this classical (i.e. non-quantum) mechanics, we assume that within the domain of space-time we are considering, analytic continuation allows us to treat $c$ as an analytic non-cyclical variable. This is consistent with the seventh comment in Section II of Ref. \cite{ZM1} identifying analytic continuation as a condition for discussing classical theoretical formalisms. Thus, for this article, derivatives can be defined in terms of the clock $c$ rather than $\psi$.

Introducing physical quantities: 
\begin{itemize}
\item $mass$ is the rate of change of the state of the system with respect to clock time, 
\begin{equation}
m  = \partial_{\hat \tau} c
\end{equation}
\item $momentum$, $p$ or $k$, is the rate of state change of $c$ with respect to an environment degree of freedom, $q$ (e.g. $q=x$ above). 
\begin{equation}
p = -  \partial_{\hat q} c
\end{equation}
\item $energy$, $\omega$, is the rate of change of a system state, $c$, with respect to observer time, $t$, 
\begin{equation}
\omega  = d_t c,
\end{equation}
where the full derivative with respect to time is defined to be the path derivative in the time direction
\begin{equation}
d_t = \partial_{\hat t}.
\end{equation}
\end{itemize}

In the first paper we considered uniform variations in the clock state with observer-extrapolated coordinate $q$. In this paper we consider the possibility of a more general spatial variation in clock coordinates that imply a spatial variation in momentum. In classical physics we consider momentum to depend on time rather than on space. Here this time dependence arises from the spatial dependence of the momentum (similar to the quantum concept of momentum as a spatial operator) and the non-orthogonality of space and time. 

To identify the notion of a trajectory, which is central to classical mechanics, we consider the origin of the space axis to follow the direction of time. By definition, this is the direction of the increase of time, and a reference location changes its coordinate due to the displacement of the origin of the space axis. We include the possibility that $p$ varies as we trace the origin of the space axis over time due to variation in the slope of the direction of time. In this way, the direction of time defines a trajectory through space-time. 

The trajectory can be written in terms of the angle of the time direction or, equivalently, in terms of the velocity, which is determined by the angle of time at a particular location, 
\begin{equation}
\begin{array}{ll}
\dot q & = k/\omega \\
&= \sin(\zeta),
\end{array}
\end{equation}
where $\zeta$ is the angle of the direction of time measured from the $\hat \tau$ axis rotated in the direction of $-\hat x$. This expression is valid even when the angle of time varies with the coordinate $q$, if the angle is taken to be the angle at the origin of coordinates.
Specifically, the trajectory is given by the time dependence of the coordinate $q$ which, due to the dynamics of the origin of coordinates, is perhaps more clearly written explicitly as measured with respect to the coordinate origin
\begin{equation}
q \rightarrow q-\bar q
\end{equation}
where $q$ is the reference location, and $\bar q$ is the origin of coordinates (normally taken to be zero). The velocity is the derivative of the trajectory
\begin{equation}
\dot q = d_t (q - \bar q),
\end{equation}
or equivalently,
\begin{equation}
\begin{array}{ll}
\dot q &=d_t  (q_0 - \bar q_0)\\
& =  - d_t \bar q_0
\end{array}
\end{equation}
where $q_0$ is the unchanging reference location and $\bar q_0$ is the displacement along the $\hat q$ direction of the origin of coordinates. The latter expression is, by geometry, the angle of time $\sin(\zeta)=k/\omega$ at the origin of coordinates.

The non-orthogonality of space and time allows us to write the time dependence of the clock in terms of its spatial dependence, i.e., the energy in terms of the momentum, which is the Hamiltonian. Similar to Eq. (\ref{generaltime}), the generalized function obtained from adding the variation along each of the coordinate axes: 
\begin{equation}
\begin{array}{ll}
H &= \partial _{\hat q} c d_t q_0 + \partial _{\hat \tau}  c d_t \tau_0  \\
&= p \dot q + m d_t  \tau_0
\end{array}
\end{equation}
is valid for arbitrary direction of time. However, we can follow the derivation of the direction of time in the introduction to obtain the Hamiltonian as a function only of momentum by maximizing $H$ with respect to the angle of time, $\zeta$, where 
\begin{equation}
\dot q = \sin (\zeta ),
\end{equation}
and 
\begin{equation}
d _t \tau_0  = \cos (\zeta ).
\end{equation}
Substituting gives the generalized function
\begin{equation}
H(p,\dot q) = p\dot q + m\sqrt {1 - \dot q^2 }.
\end{equation}
Maximizing $H(p,\dot q)$ with respect to $\dot q$ by setting
\begin{equation}
\partial H/\partial \dot q = 0
\label{maximize}
\end{equation}
gives 
\begin{equation}
p = m\dot q/\sqrt {1 - \dot q^2 }, 
\end{equation}
solving for $\dot q$ and inserting the result into $H(p,\dot q)$ gives 
\begin{equation}
H(p) = \sqrt {p^2  + m^2 }. 
\end{equation}
which is the Hamiltonian as a function only of momentum.

The momentum and velocity both describe the relationship between the system and environment. $p$ plays the role of an extrinsic parameter and $\dot q$ is determined by it, and thus can be considered a function $\dot q(p)$. Considering $p$ to be variable we can therefore write  
\begin{equation}
\begin{array}{ll}
 dH(p)/dp &= \partial H(p,\dot q)/\partial p + (\partial H(p,\dot q)/\partial \dot q)(d\dot q/dp) \\
 &= \partial H(p,\dot q)/\partial p \\
 &= \dot q,
\end{array}
\end{equation}
where the first equality is the chain rule, the second equality arises from the variational time direction determination as given by 
Eq.(\ref{maximize}), and the final equality from the explicit form of $H(p,\dot q)$. This is the first of Hamilton's equations.

If $p$ varies along the trajectory $q(t)$ (i.e. recall that $q=q-\bar q=q_0-\bar q_0(t)$), then $\dot q(p)$ and $H(p)$, are implicit functions of $q$. The time derivative of $p$ is given by 
\begin{equation}
\begin{array}{ll}
dp/dt &= (dp/dq)(dq/dt) \\
&= (dp/dq)\dot q \\
&= (dp/dq)(dH(p)/dp) \\
&= dH(p)/dq, 
\end{array}
\label{dpdtfirst}
\end{equation}
where the third equality follows from Hamilton's first equation. This is a positional change in the observed `kinetic energy' that occurs along the trajectory. The change of $p$ can be considered as extrinsic to the system. Assuming causation we postulate an extrinsic cause, `The Force.' Specifically, we assume a conserved `total energy' 
\begin{equation}
H(p,q) = H(p) + \phi (q), 
\label{spaceterminH}
\end{equation}
which is the combined rate of state change of the system and environment (or system and other system that is also separate from the environment). Conservation implies 
\begin{equation}
dH(p,q)/dq=0,
\end{equation}
or 
\begin{equation}
d\phi(q) /dq =  - dH(p) / dq, 
\label{phidefine}
\end{equation}
and inserting into Eq. (\ref{dpdtfirst}),
\begin{equation}
\begin{array}{ll}
dp / dt &=  - d\phi(q) / dq \\
&=  - \partial H(p,q) / \partial q, 
\label{Hamiltonssecond}
\end{array}
\end{equation}
which is the second of Hamilton's equations. 

\section{Lagrange's equations of motion}

To discuss the Lagrangian formalism we define 
\begin{equation}
L =  - m d _t \tau_0, 
\end{equation}
so that 
\begin{equation}
H = p\dot q - L.
\label{Ldefine}
\end{equation}
The explicit form of $H(p,\dot q)$ implies that $L$ does not depend on $p$ directly, but the determination of the direction in time is equivalent to 
\begin{equation}
dL(\dot q)/d\dot q = p. 
\label{Lequation1}
\end{equation}
Lagrange's equations of motion follow from defining the augmented Lagrangian from
\begin{equation}
H(p,q) = p\dot q -  L(q,\dot q)
\end{equation}
where $H(p,q)$ is given by Eq. (\ref{spaceterminH}) so that
\begin{equation}
L(q,\dot q) = L(\dot q) - \phi (q)
\end{equation}
With the definition of $L(q,\dot q)$  we revise Eq. (\ref{Lequation1}) by using a partial derivative:
\begin{equation}
\partial L(q,\dot q)/ \partial \dot q = p. 
\end{equation}
Taking the derivative with respect to time of this equation and inserting Hamilton's second equation, Eq. (\ref{Hamiltonssecond}), we obtain
\begin{equation}
(d/dt) \partial L(\dot q,q)/\partial \dot q = \partial L(\dot q,q)/\partial q. 
\end{equation}
which is Lagrange's equation of motion. 

\section{Comments}

We see that the dynamical equations of classical mechanics can be inferred from orienting the direction of time toward the maximum rate of change of the clock, where the non-orthogonality of space and time plays a role in the transformations. There are a number of comments that follow from the derivation of Hamilton's and Lagrange's equations of motion. 

First, it is helpful to compare the central assumption of ZM theory (orienting the direction of time toward the maximum rate of change of the clock) as a variational principle, with Hamilton's principle, and other conventional variational principles. Each obtains the actual trajectory as an extremum over a set of hypothetical trajectories. Hamilton's principle states \cite{Goldstein} that the action,
\begin{equation}
S = \int {L(q,\dot q) dt} ,
\end{equation}
is an extremum along the correct trajectory when varying over all trajectories with fixed end points. A modified version of this principle allows the direct derivation of Hamilton's equations by variation of the integral
\begin{equation}
\int {(p\dot q - H(q,p)) dt} ,
\end{equation}
treating $q$ and $p$ as independent variables. In both of these conventional variational principles, the potential energy is assumed to be fixed. By contrast, in ZM theory, the spatially dependent momentum $p(q)$ is assumed to be fixed, the velocity $\dot q$ is allowed to vary, and the (gradient of the) potential is defined after the determination of the trajectory by Eq. (\ref{phidefine}). For the one dimension discussed here, since all of these principles give rise to the same trajectories, as given by either Hamilton's or Lagrange's equations, they are equivalent upon transformation between sets of independent variables. It has been argued that the only fundamental justification for Hamilton's principle is that it generates Lagrange's equations of motion.\cite{Goldstein} Thus, ZM theory has, at this level of discussion, equivalent justification.

Second, we note that the mathematical expressions for the velocity, $v$, that appeared in the discussion of Lorentz covariance in Ref.  \cite{ZM1}, and that used here, $\dot q$, are equivalent.

Third, since in ZM theory the value of $k$ is related to the field variable at a given location, it might be asked how there could be different velocities at the same location separated in time. This would happen, for example, in oscillatory motion. The answer to this question,  to be discussed in subsequent papers, has to do with multiple dimensions, which can also be described as the existence of a multisheeted space.

Fourth, as in classical mechanics and quantum mechanics, the existence of a potential $\phi(q)$ is assumed as an effective field to account for system behavior where causality requires environment coupling. This then leads to Newton's Second law. The assumption of causality is then validated by consistency in describing $\phi(q)$ in terms of environmental attributes (e.g. coordinates of other systems that are subsystems of the environment). This has not been shown in this formalism thus far. 

Fifth, the treatment in this paper considers only a single spatial dimension. Otherwise, additional angles describing the direction of time are necessary. 
With sufficient simplifying assumptions that dynamically decouple the dimensions, including taking the small angle (non-relativistic) limit, a simple multi-dimensional formalism is possible. This could correspond to the usual generalization in the context of Hamilton's and Langrange's formalisms. Without such simplifications the dimensions are not intrinsically independent and for example, symmetry operators such as rotations can transform one dimension into another. Therefore, multi-dimensional generalizations are non-trivial. Their discussion is deferred.

Sixth, in the conventional Hamiltonian and Lagrangian formalism there is no fundamental reason that the potential energy is not a function of velocity, whereas here this can be inferred. However, this conclusion is valid in the one dimensional case treated in this paper, but not in higher dimensions. This can be seen in the electromagnetic field tensor which only has an electric field in 1+1 dimensions, and velocity coupling arises in higher dimensions in the context of spatial rotation. These expressions, and conclusions, are therefore inadequate in higher dimensions where spatial rotations couple to velocity and Dirac's equation is necessary. Subsequent papers will consider three spatial dimensions, including  the Dirac equation and electrodynamics.

Seventh, it is interesting to note that the derivation of a classical trajectory in this paper relies upon the dynamics of the origin of coordinates rather than the field behavior at the reference location.  This suggests the interpretation that velocity and acceleration are to be understood as those of the observer. Since velocity and acceleration are always relative, this is surely not a conceptual or formal problem. It is reasonable to expect that an alternative equivalent representation would lead to changes in the angle of time at the reference location as a function of time, resulting in an explicit dynamics at this location. We do not pursue this approach here. Still it is interesting to note that in the derivation in this paper the potential energy explicitly arises because of the effect of the angle of time at locations that are not at the reference location. The potential energy therefore is due to non-local effects, as it is in classical source equations.

\end{document}